# L'entreprise franco-roumaine face au Internet

**Assist.Prof. Diana Sophia Codaţ, Ph.D.Candidate**
„Tibiscus" University of Timişoara, România

**Rezumat:** Obiectivul principal al prezentului articol este analiza activitatii intreprinderilor franceze din Romania ca urmare a utilizarii tot mai acute in sfera productiva a tehnologiilor de informatie si comunicatie. Ansambul convergent de tehnologii de informatie si comunicatie, procesul de mondializare economica au dus la o vasta transformare a activitatii economice. Articolul face parte dintr-o serie de studii efectuate in anii 2007 si 2008 asupra firmelor franceze implantate in România.

L'objectif prioritaire de la présente étude est l'analyse de l'activité de l'entreprise française en Roumanie suite à l'irruption dans la sphère productive des technologies de l'information et la communication. Ces transformations, qui sont traduites dans des changements remarquables dans les deux inputs de base de l'activité de l'entreprise (le capital et le travail), ainsi que dans les pratiques de l'entreprise et dans l'élément déterminant de la croissance à long terme - l'innovation -, devraient empiriquement être contrastées à partir d'une hypothèse générale.

L'ensemble convergent des TIC, le processus de mondialisation économique et les changements ont donné lieu à une vaste transformation de l'activité économique, que nous groupons sous le concept de nouvelle économie (économie de la connaissance) et qui, depuis le versant de l'analyse économique peut être abordée tant du point de vue du cycle économique (macroéconomie) comme du point de vue du marché, c'est-à-dire, de l'interaction entre les agents économiques (micro-économie).

Dans le cadre de l'utilisation et la pénétration d'Internet dans les organisations de l'entreprise on peut initialement distinguer trois grandes étapes :

1ª. étape de base : Se réfère à l'utilisation d'Internet comme système de recherche d'information et de courrier électronique.

2ª. étape intermédiaire : Liée l'application d'Internet comme canal de vente (e-commerce) et instrument de base de relation avec des collaborateurs (fournisseurs et employés) pour la réalisation tâches, de tant de consultation et d'information, comme administratives.

3ª. étape avancée : Il implique l'utilisation d'Internet pour la réalisation, en outre, d'activités comme l'e-recrutement (sélection de personnel à travers Internet), e-learning (formation à travers Internet), utilisation de téléphonie IP, etc., comme canal d'intégration dans la stratégie et les processus patronaux.

L'intensification de son utilisation dans les entreprises est un facteur fondamental pour le décollage compétitif de ces dernières, puisque, entre autres, elle offre les avantages suivants :
- Introduction sur un nouveau marché potentiel.
- Obtention et diffusion de tout type d'information
- Accès à des marchés internationaux.
- Service total (24 heures au 24/365 jours par an).
- Information immédiate sur les changements et les nouveautés.
- Communication interactive

À ce diagnostic il est nécessaire d'ajouter les dangers dans l'utilisation d'Internet et qui donnent lieu encore à une plus basse utilisation de ce dernier, comme par exemple:
- Communication lente ou instable.
- Perdre de temps par navigation insignifiante.
- Excessive complication technique.
- Risque élevé de virus ou pirates avec accès à information

Bien que le niveau d'accès à Internet soit très important dans toutes les entreprises franco-roumaines - 90.9% disposait de connexion à Internet en printemps 2007- les entreprises du secteur de l'industrie de l'information et les entreprises des services intensifs en connaissance présentent des pourcentages significativement supérieurs. Ainsi, 98.4% et 98.3% des entreprises de chacun des deux secteurs a une connexion à Internet. Pour sa part, l'industrie de technologie moyenne, avec seulement 79.7% des entreprises reliées à Internet, présente un pourcentage inférieur au reste de secteurs patronaux. Comme dans le cas des ordinateurs, la taille de l'entreprise il est indépendant du fait de disposer ou non de la connexion à Internet. Bien que l'analyse des données laisse entrevoir une certaine corrélation entre la dimension et le fait d'avoir ou non Internet (au fur et à mesure qu'augmente la taille de l'entreprise, le pourcentage d'entreprises qui sont reliées à Internet est supérieur), on n'apprécie pas de différences statistiquement significatives.

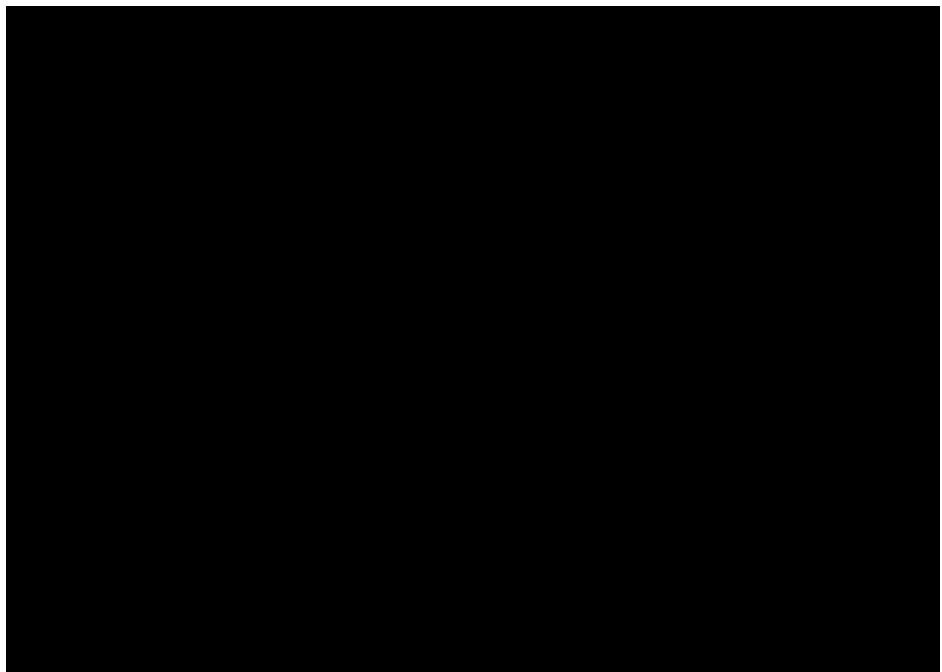

*Figure no 1 : Type de connexion à Internet Dans des pourcentages sur le total d'entreprises franco-roumaines*

Pratiquement 44% d'entreprises franco-roumaines est reliée à Internet à travers une ligne ADSL. Le pourcentage d'entreprises franco-roumaines en rapport avec d'autres systèmes à bandes larges est très sous. Ainsi, seulement 9% le fait à travers le câble et, comme il se détache des données obtenues, l'utilisation de satellite pour être relié à Internet est que de 2%. L'industrie de technologie faible et moyenne est celle qui présente une tendance supérieure à utiliser le modem (ou réseau de téléphonie de base) (70%), en tenant compte du fait que ce type de connexion seul est choisi par 45% des entreprises franco-roumaines. Quant à d'autres types de connexion, il convient de commenter que l'ISDN est le troisième système plus utilisé (16%) après l'ADSL et le modem. Tandis que dans le cas de disposer ou non d'ordinateurs et de connexion à Internet on donne une indépendance claire des données en ce qui concerne la taille des entreprises, dans le cas de la disposition de réseaux locaux (LAN), reliés ou non à d'autres réseaux publics ou privés (WAN), on met en évidence une dépendance claire entre ces deux ampleurs. Les entreprises avec plus de 20 travailleurs tendent à utiliser plus les réseaux locaux (entre 53%, celles de 20 à 99 travailleurs et 72% celles de plus de 100 travailleurs) que de petites entreprises, avec moins de 20 travailleurs (entre 23%, celles de 6 à 9 travailleurs et 28%, celles de 10 à 19 travailleurs). La nécessité de disposer de communication interne et de partager une information entre un nombre important de travailleurs fait que la présence de ces technologies est plus importante dans les grandes entreprises.

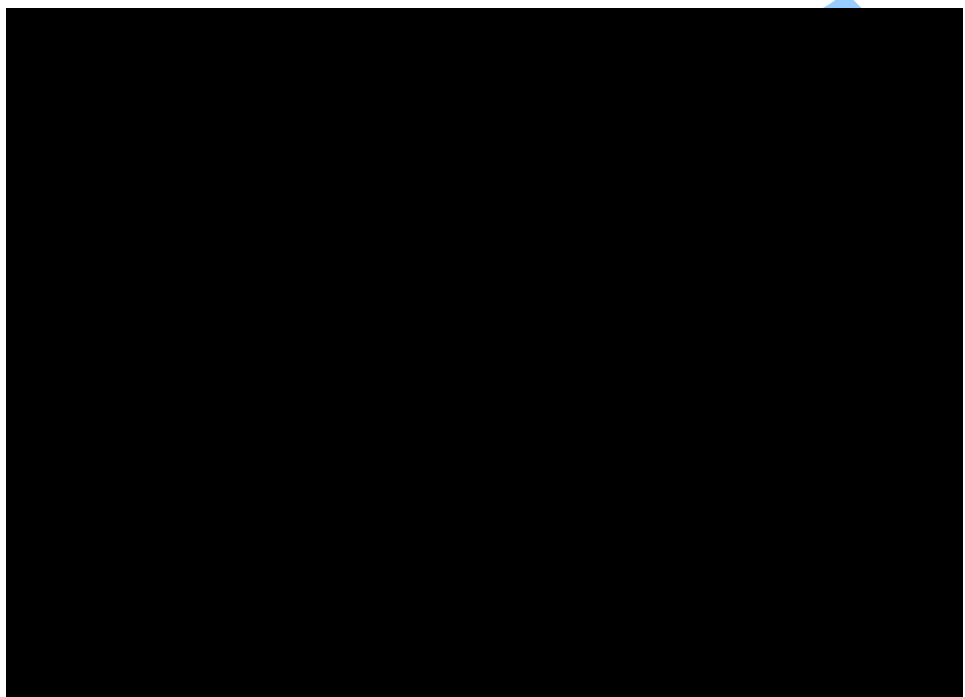

*Figure no 2: Réseau locale dans les entreprises en pourcentages sur le total d'entreprises franco-roumaines*

L'utilisation interne des TIC, les réseaux locaux constituent un outil fondamental, puisqu'ils facilitent l'organisation en réseau de l'activité de l'entreprise. Plus de la moitié des entreprises franco-roumaines (53%) dispose de réseaux locaux.

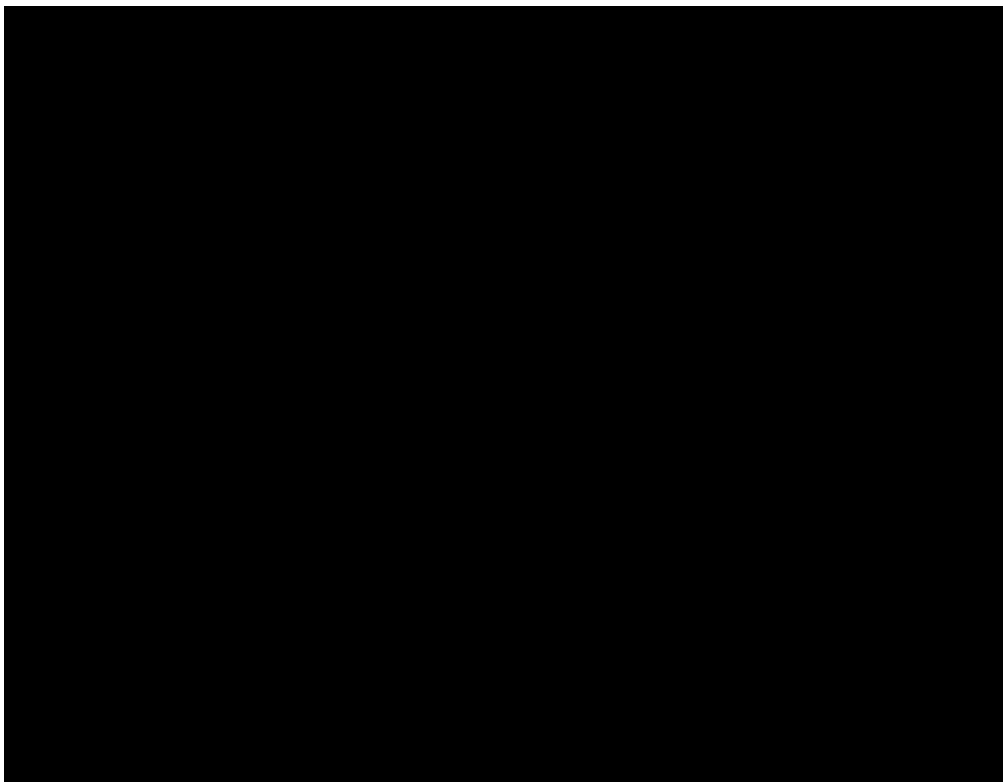

*Figure no 3: L'Utilisation des systèmes d'échange électronique de données avec des fournisseurs et des clients.*

Bien que le réseau local constitue un important élément pour l'organisation en réseau des entreprises au niveau interne, les systèmes d'échange électronique de données sont pour le niveau externe. Évidemment, l'entreprise réseau a besoin qu'il existe une importante connexion entre des producteurs, consommateurs et fournisseurs. En ce sens, il convient de souligner que les systèmes d'échange électronique de données facilitent l'intégration stratégique des fournisseurs et des clients dans l'organisation et, en outre, consolident une vision globale de toutes les ressources utilisées pour la réalisation de défis et objectifs partagés. Ce lien stratégique permet le développement de synergies pour aborder des projets communs d'une plus grande complexité, condition indispensable pour l'adaptation de l'activité patronale à une demande à caractère globale et dans évolution constante. La même manière que le pourcentage d'entreprises avec des réseaux locaux augmente avec le nombre de travailleurs, nous pouvons aussi trouver une relation semblable dans le cas du pourcentage d'entreprises qui utilise des systèmes d'échange électronique de données avec des fournisseurs et des clients.

Tandis que seulement 44% des petites entreprises franco-roumaines dispose de ces systèmes, dans le cas des entreprises moyennes et grandes ce pourcentage devient de 84% et de respectivement 91%.

Si nous analysons cette variable par des secteurs d'activité, l'industrie de faible technologie, moyenne et haute, autour de 20% des entreprises de ces trois branches d'activité applique ces systèmes dans son organisation, face à 36% des entreprises de l'industrie de l'information. Nous avons, par conséquent, une relation directe entre la disposition de ce type d'équipement et l'orientation en réseau de l'entreprise elle-même, tant en ce qui concerne leur dimension comme en ce qui concerne leur activité. Pratiquement, la moitié des entreprises franco-roumaines dispose de page web propre (56.1%). Ce pourcentage est aussi celui qu'il caractérise, par le poids qui représente en ce qui concerne le reste des secteurs, aux entreprises de services moins intensifs en connaissance (48.2%). Bien qu'il y ait de petites différences significatives entre des secteurs patronaux, il convient de souligner qu'autour de 60% des entreprises franco-roumaines de l'industrie de l'information et de l'industrie de haute technologie (70% et 65%, respectivement) ils sont présents dans Internet à travers leur propre page web.

**Conclusions:**

En synthèse, la pénétration des équipements numériques, de l'Internet et du courrier électronique, dans l'entreprise franco-roumaine ont présenté surtout une tendance marquée à la hausse durant les la dernière période, tendance dans laquelle doivent aussi s'inclure les entreprises de dimensions plus petites. Cependant, il convient de souligner l'évolution expansive des pages web et du commerce électronique, bien que ce dernier ait encore une présence limitée dans l'activité organisationnelle.